\documentstyle[referee,psfig]{mn} 
 
\newif\ifAMStwofonts 
 
\def\kms{\thinspace\hbox{$\hbox{km}\thinspace\hbox{s}^{-1}$ }}
\def\ergs{\thinspace\hbox{$\hbox{erg}\thinspace\hbox{s}^{-1}$}}
\def\percc{\thinspace\hbox{$\hbox{cm}^{-3}$}}
\def\ha{\hbox{$\hbox{H}\alpha$ }}
\def\hb{\hbox{$\hbox{H}\beta$ }}
\def\msun{\thinspace\hbox{$\hbox{M}_{\odot}$}}
\def\etal   {{\sl et\nobreak\ al.\ }}
\def\ni{\noindent}                                       

\def\si{$\sim$ }                        
\def\sol{$_\odot$ }                     

\newcommand{\E}[1]{\,10^{#1}}                         
\newcommand{\ET}[1]{\times 10^{#1}}             
\newcommand{\zb}[1]{\left[ {#1} \right]}           
\newcommand{\zp}[1]{\left( {#1} \right)}               

\ifoldfss 
  \ifCUPmtlplainloaded \else 
    \NewTextAlphabet{textbfit} {cmbxti10} {} 
    \NewTextAlphabet{textbfss} {cmssbx10} {} 
    \NewMathAlphabet{mathbfit} {cmbxti10} {} 
    \NewMathAlphabet{mathbfss} {cmssbx10} {} 
  \fi 
  \ifAMStwofonts 
    \ifCUPmtlplainloaded \else 
      \NewSymbolFont{upmath} {eurm10} 
      \NewSymbolFont{AMSa} {msam10} 
      \NewMathSymbol{\upi}     {0}{upmath}{19} 
      \NewMathSymbol{\umu}     {0}{upmath}{16} 
      \NewMathSymbol{\upartial}{0}{upmath}{40} 
      \NewMathSymbol{\leqslant}{3}{AMSa}{36} 
      \NewMathSymbol{\geqslant}{3}{AMSa}{3E}

      \let\geq=\geqslant \let\ge=\geqslant 
    \fi 
  \fi 
\fi 
 
\ifnfssone 
  \newmathalphabet{\mathit} 
  \addtoversion{normal}{\mathit}{cmr}{m}{it} 
  \addtoversion{bold}{\mathit}{cmr}{bx}{it} 
  \newmathalphabet{\mathbfit} 
  \addtoversion{normal}{\mathbfit}{cmr}{bx}{it} 
  \addtoversion{bold}{\mathbfit}{cmr}{bx}{it} 
  \newmathalphabet{\mathbfss} 
  \addtoversion{normal}{\mathbfss}{cmss}{bx}{n} 
  \addtoversion{bold}{\mathbfss}{cmss}{bx}{n} 
  \ifAMStwofonts 
    \ifCUPmtlplainloaded \else 
      \UseAMStwoboldmath 
      \makeatletter 
      \new@mathgroup\upmath@group 
      \define@mathgroup\mv@normal\upmath@group{eur}{m}{n} 
      \define@mathgroup\mv@bold\upmath@group{eur}{b}{n} 
      \edef\UPM{\hexnumber\upmath@group} 
      \new@mathgroup\amsa@group 
      \define@mathgroup\mv@normal\amsa@group{msa}{m}{n} 
      \define@mathgroup\mv@bold\amsa@group{msa}{m}{n} 
      \edef\AMSa{\hexnumber\amsa@group} 
      \makeatother 
      \mathchardef\upi="0\UPM19 
      \mathchardef\umu="0\UPM16 
      \mathchardef\upartial="0\UPM40 
      \mathchardef\leqslant="3\AMSa36 
      \mathchardef\geqslant="3\AMSa3E 

      \let\geq=\geqslant \let\ge=\geqslant 
    \fi 
  \fi 
\fi 
 
\ifnfsstwo 
  \DeclareMathAlphabet{\mathbfit}{OT1}{cmr}{bx}{it} 
  \SetMathAlphabet\mathbfit{bold}{OT1}{cmr}{bx}{it} 
  \DeclareMathAlphabet{\mathbfss}{OT1}{cmss}{bx}{n} 
  \SetMathAlphabet\mathbfss{bold}{OT1}{cmss}{bx}{n} 
  \ifAMStwofonts 
    \ifCUPmtlplainloaded \else 
      \DeclareSymbolFont{UPM}{U}{eur}{m}{n} 
      \SetSymbolFont{UPM}{bold}{U}{eur}{b}{n} 
      \DeclareSymbolFont{AMSa}{U}{msa}{m}{n} 
      \DeclareMathSymbol{\upi}{0}{UPM}{"19} 
      \DeclareMathSymbol{\umu}{0}{UPM}{"16} 
      \DeclareMathSymbol{\upartial}{0}{UPM}{"40} 
      \DeclareMathSymbol{\leqslant}{3}{AMSa}{"36} 
      \DeclareMathSymbol{\geqslant}{3}{AMSa}{"3E} 

      \let\geq=\geqslant \let\ge=\geqslant 
    \fi 
  \fi 
\fi 
 
\ifCUPmtlplainloaded \else 
  \ifAMStwofonts \else 
    \def\upi{\pi} 
    \def\umu{\mu} 
    \def\upartial{\partial} 
  \fi 
\fi 
 
\title[SN~1997ab]
{The circumstellar medium of the peculiar supernova SN1997ab.\thanks{Based on 
observations made with the 4.2-m WHT operated on the island of La Palma by the
Isaac Newton Group in the Spanish Observatorio del Roque de los Muchachos of the
Instituto de Astrof\'{\i}sica de Canarias.}}

\author[Salamanca et al.]
        {Isabel Salamanca$^{1,2}$, Roberto Cid-Fernandes$^3$, Guillermo Tenorio-Tagle$^{2,4,5}$,  
     \newauthor Eduardo Telles$^6$, Roberto J. Terlevich$^2$ and Casiana Mu\~noz-Tu\~n\'on$^7$  \\
$^1$Observatory of Leiden, Postbus 9513, NL-2300 RA Leiden, The Netherlands.\\
$^2$Royal Greenwich Observatory, Madingley Road, Cambridge CB3 OEZ, U.K. \\
$^3$Dep. de f\'{\i}sica, CFM, UFSC, Campus Universit\'ario, Trinidade, Caixa Postal 
476, 88040-900 Florian\'opolis, SC, Brasil.\\
$^4$Institute of Astronomy, Univ. of Cambridge, Madingley Road, Cambridge CB3 0HA, U.K. \\
$^5$Instituto Nacional de Astrof\'{\i}sica Optica y Electronica, AP 51, 72000 Puebla,
M\'exico. \\
$^6$Observatorio Nacional, Departamento de Astrof\'{\i}sica, Rua Jose Cristino 77,
20921-400 - Rio de Janeiro, Brasil. \\
$^7$Instituto de Astrof\'{\i}sica de Canarias, E-38200 La Laguna, Tenerife, Spain. \\
}

\date{Accepted ...
      Received ...; 
      in original form ...} 
 
\pagerange{\pageref{firstpage}--\pageref{lastpage}} 
\pubyear{1998} 

\begin{document}
\maketitle 
 
\label{firstpage} 

\begin{abstract}
We report the detection of the slow moving wind into which    
the compact supernova remnant SN~1997ab is expanding. 
Echelle spectroscopy provides 
clear  evidence for a wellresolved narrow (Full Width at Zero Intensity, FWZI \si 180 \kms) P-Cygni 
profile, both in \ha and 
\hb, superimposed
on the broad emission lines of this compact supernova remnant.
From theoretical arguments we know that
the broad and strong emission lines imply a circumstellar density ($n 
\geq 10^7$ \percc). This, together with our detection, implies a massive and slow
stellar wind experienced by the progenitor star shortly prior to 
the explosion.
\end{abstract}



\section{Introduction}

SN1997ab, like SN1987F or SN1988Z (Filippenko 1989, Turatto et al. 1993) 
is a peculiar type II supernova 
(SN) with a luminosity of $M_B = -17.5$ {\it one year after its discovery} 
(Hagen, Engels \& Reimers 1997). 
It was first detected on April 11, 1996, in the dwarf galaxy HS 0948+2018, far from the
galaxy center, showing broad Balmer emission lines, FWHM \si 2500 to 3000 \kms, without the
typical P-Cygni profiles of type II SN. The redshift derived from the narrow emission 
lines seen in SN 1997ab is 0.012 (Hagen et al. 1997). These
peculiar SN, also called ``Seyfert~1 impostors"
(Filippenko 1989), given their profound resemblance to the 
emission spectra from
Seyfert~1 nuclei, are sometimes also identified as radio supernovae (RSN; see Van Dyk et al. 
1993, 1996). 
The standard explanation for these sources (Chugai 1990, Terlevich et al. 1992, 1995)
demands the expansion of a supernova shock into a dense circumstellar medium,
and thus it is believed that what we see is the SN remnant and not the 
SN itself. Indeed,
the release of $10^{51}$ erg into a high density medium ($\geq 10^6$ \percc)
is known to lead 
to a rapidly evolving remnant, which skips its quasi-adiabatic
Sedov evolutionary phase to
become strongly radiative, reaching luminosities $\geq$ 10$^9$ L\sol 
even before the ejecta is fully thermalized at the reverse shock 
(see Wheeler \etal 1980; Terlevich et al. 1992). 
The rapid evolution  is a direct consequence  of  the high densities,
which promote the earlier 
onset of strong radiative cooling in the matter swept up 
 and strongly accelerated by the blast wave.
Strong cooling causes the collapse of the swept up gas into
a thin shell while still moving at several thousands of \kms~and with it, to
an outburst of radiation capable of ionizing the cooling shell,
the unshocked ejecta and the background gas. In this way 
the supernova energy, the 10$^{51}$ erg originally released as kinetic energy,
is radiated away in just a matter of a few years. This causes
the most luminous SN remnants, although they only
acquire sizes \si a few 10$^{16}$ cm, and hence have been termed ``compact'' 
supernova remnants (CSNRs).  

Here we provide unambiguous evidence for the existence of a slowly expanding, 
dense circumstellar
medium around SN~1997ab and  work out its relevant physical properties.

\section{The observations}

On May 30 1997, the supernova remnant, SN1997ab (Hagen et al. 1997), 
was observed with the Utrecht Echelle
spectrograph at the 4.2-m William Herschel Telescope at the
``Roque de los Muchachos'' observatory (La Palma, Spain).
A 1200s exposure of the supernova remnant was obtained using a 2148$\times$2148    Tek CCD
and the 79.0 lines/mm echelle, reaching a S/N $\sim$ 30 at H$\alpha$.
The orders covered are 45 to 30, from
$\sim$ 4130 \AA - 7340 \AA, including \hb and \ha.
The spectral resolution of our setup was 0.13 \AA~in the \hb region and 0.18 \AA\ 
in the \ha
region. This led to a resolution of $\sim$ 8 \kms.
The CCD frames were reduced using standard procedures with the tasks
in IRAF.  The data was debiased, trimmed and flat-field using a
normalized flat field produced with apflatten.  The spectra were
extracted using doecslit with appropriate parameters.  Wavelength
calibration was performed by using the comparison spectrum of
Thorium-Argon arcs. Flux
calibration was accomplished through the observation of standard stars. 
The atmospheric extinction
correction was also applied at the time of calibration using the mean
extinction curve for La Palma.
No redshift correction was applied, and therefore the wavelengths are the
observed ones, unless otherwise stated.
The seeing was less than 1.5 arc seconds.

\subsection{Line analysis}

\begin{figure*}
\psfig{figure=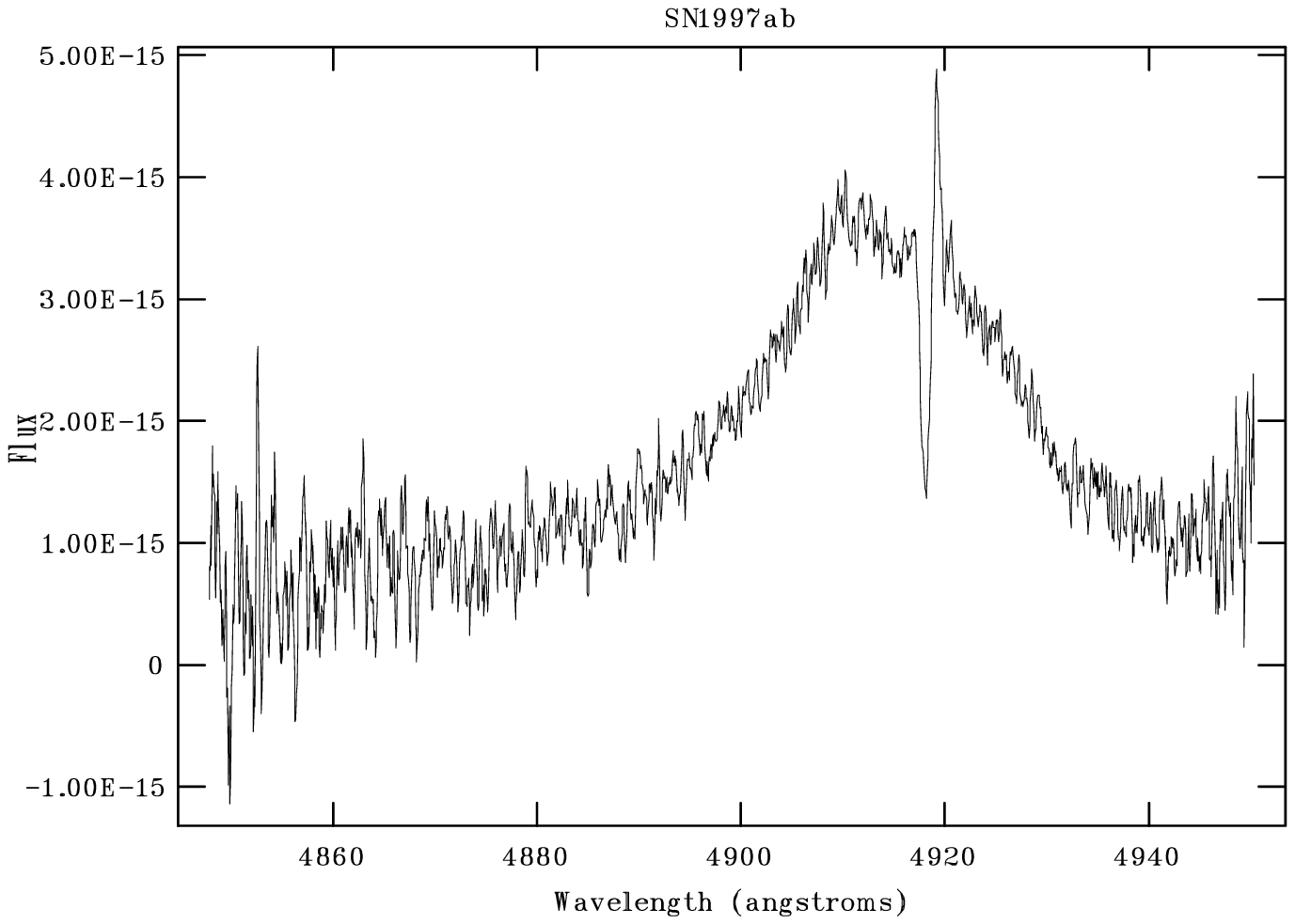} 
\psfig{figure=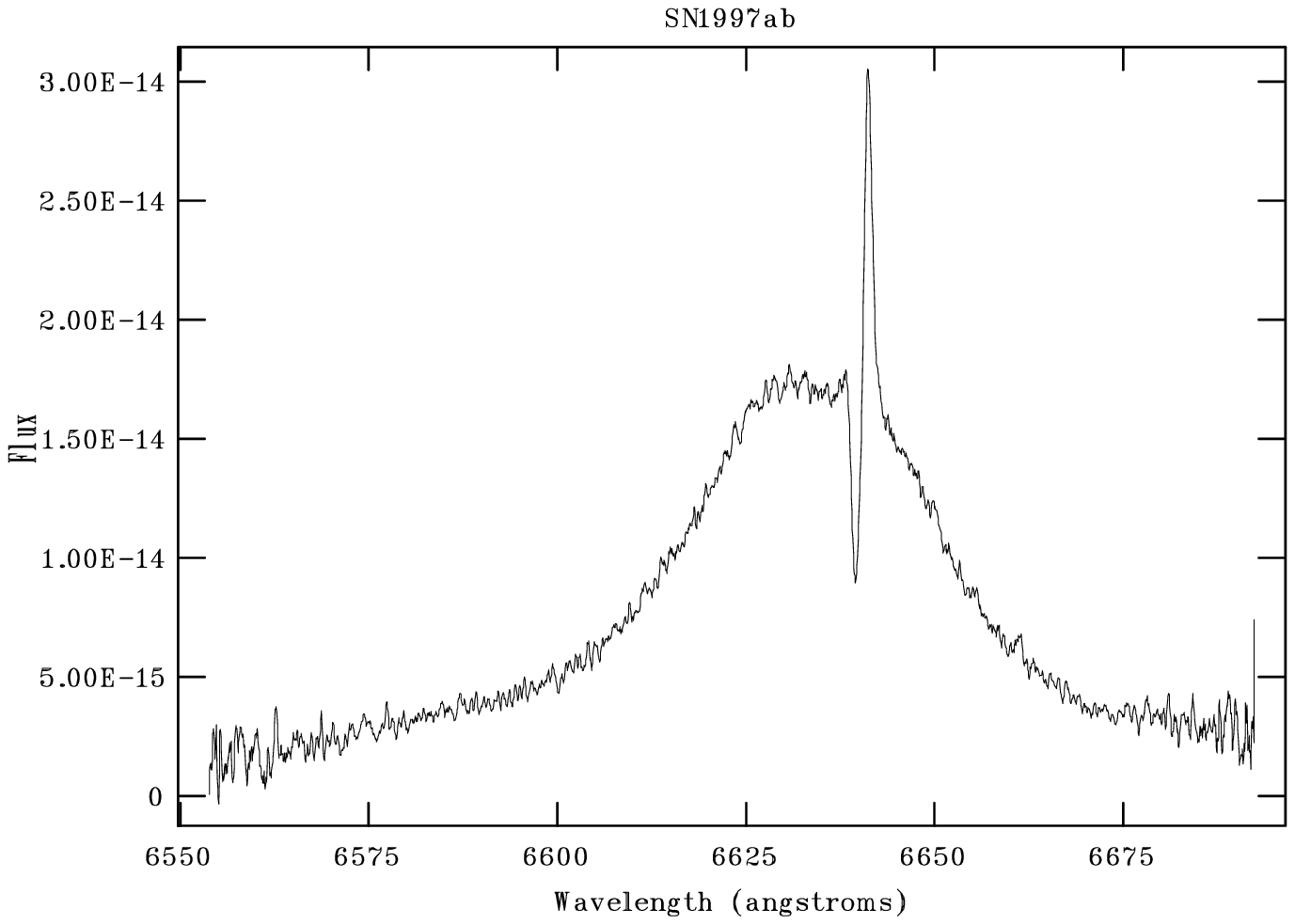} 
 \caption{The \hb and \ha profiles of SN1997ab. The flux is in units of erg \AA$^{-1}$ s$^{-1}$ cm$^{-2}$. A ``smooth'' has been applied to both spectra to reduce the noise in the wings. }
\label{hahb}
\end{figure*}

Figure~\ref{hahb} shows the spectral regions around \ha and \hb lines. 
Both of these
lines have two components: a broad component with a 
FWHM $\sim$ 1800 \kms~and a narrow (FWZI $\sim$ 180 \kms) P-Cygni profile 
superimposed on the broad lines.
Note that the broad components present a flat-topped profile,
typical of expanding shells, and that the P-Cygni narrow lines are
somewhat displaced towards longer wavelengths. 
The broad lines indicate a large expansion  velocity of (at least) \si 2000 to 4000 \kms,
as measured from their red and blue wings, respectively.
However, the fact that the P-Cygni profiles are shifted implies that
the
red part of the broad components of both, \ha and \hb, are probably self-absorbed
and therefore missing. As a consequence, the 
intrinsic true width of the broad profiles 
could be almost twice as large as what
one can at first glance measure.  Also, given
the small wavelength range covered in each portion of our spectra, 
the broad \ha and \hb lines are not fully sampled 
and their wings and
adjacent continuum are missing. Therefore, regardless of their asymmetry, 
the FWZI cannot be measured in our Echelle spectra.
A better determination  of the FWZI is to be found in 
previous published (low-dispersion) spectra of this CSNR 
(Hagen et al. 1997, Salamanca et al. 1998) which
indicates a velocity of $\sim $ 6600 \kms.

The narrow P-Cygni lines shown in  Figure~\ref{hahb} on 
top of the broad \ha and \hb, must arise in a medium with a much smaller 
velocity, $v_w$ \si 90 \kms, as deduced from their half width at zero intensity.

\section{Discussion}

 The detection of narrow P-Cygni profiles in \ha and \hb,
superimposed on the broad emission lines produced by the compact 
supernova remnant SN~1997ab, proves the existence of a
slowly expanding, dense medium into which the CSNR shock evolves.
From the analysis of the broad lines one can derive the physical conditions 
of the SN ejecta and its main shock. On the other hand, the narrow P-Cygni lines 
tell us about the density and extent of the CSM.
One word of warning however: some of the narrow Balmer emission is contaminated
by the extended emission from the host galaxy, a starburst galaxy of low
excitation. Nevertheless, the emission coming from the host galaxy is very
weak, and therefore its contamination is negligible in a first approximation.
More detailed analysis will be referred to  uncoming publications (Salamanca
et al. 1998).

To estimate the density and extent of the envelope, we assumed
that the wind from the progenitor star that created it is freely expanding
with a velocity, $v_w$. The density ($\rho$) of the slow wind then falls as  r$^{-2}$, 
under the assumption of a constant mass-loss rate, \.M,
\begin{equation}
  \label{mdot}
\dot M = 4 \pi r^2 v_w \rho  
\end{equation}

\subsection{The broad lines and the CSNR shock}

When the shock in the circumstellar medium is the dominant source of
luminosity, rather than the release and escape of stored radioactive
energy, then the luminosity in the broad H$\alpha$\ line is simply
proportional  to the kinetic energy dissipated per unit time across
the shock front (Terlevich 1994, Chugai 1991), $L_s = 1/2 \dot{M}_{CSM}
v_s^2$, where $v_s$\ is the shock velocity and $\dot{M}_{CSM} =
\frac{\dot{M}}{v_w} v_s$\ is the rate at which the circumstellar medium (CSM) 
is swept up. The r$^{-2}$ density distribution will ensure that the shock
velocity can be regarded as constant (until an appreciable fraction of the
kinetic energy is radiated away). This leads to
\begin{eqnarray}
  \label{Lha}
    L^{Broad}_{H\alpha} & = & 
        \frac{1}{4} \epsilon_{H\alpha} \frac{\dot M}{v_w} v_s^3  \\
        & = &
        \pi \epsilon_{H\alpha} r^2 \rho v_s^3
\end{eqnarray}
\ni where we have taken into account the fact that only half of the 
radiation from the gas cooling behind the  shock
will contribute to the ionization of the shell, while the 
other half will be transported outwards.
The value of the efficiency factor $\epsilon_{H\alpha}$, 
depends on the time 
elapsed since the explosion of the supernova, and ranges from a maximum 
value \si 0.1 when catastrophic cooling starts, and then steadily drops to 
nearly zero within the evolutionary time of the CSNR
(Terlevich 1994 and Cid-Fernandes \& Terlevich 1994).

In the case of SN~1997ab, $v_s$ and $v_w$ are easy to determine 
from the FWZI of the broad 
and narrow
lines, respectively: $v_s$ is about \si 6600 \kms 
and for $v_w$ we shall adopt a value of 90 \kms (see previous section). 
The luminosity in \ha
is \si $ 5 \times 10^{41}$ erg s$^{-1}$. This value, taken from Hagen et al. (1997),
is probably a lower limit, since we did not take into account reddening or
obscuration effects.
The value of $\epsilon_{H\alpha}$ is directly related to the Balmer decrement, since
it depends essentially on the density of the CSM. From previous spectra
of SN~1997ab (Hagen et al. 1997) we have $\frac{H\alpha}{H\beta} \sim 6$ which led us to the 
conclusion that $\epsilon_{H\alpha}$
must be \si 0.1 (see also Terlevich 1994). Note also that, 
as  SN~1997ab was not seen on April 5th 1995 (Hagen et al. 1997), then 
an upper limit to the time elapsed
since the explosion up to the end of
May 1997, is 783 days, which is again consistent with large values of  
$\epsilon_{H\alpha}$.

The values of $v_s$, $v_w$, L$_{H\alpha}$, etc. have been obtained  
from spectra
taken at epochs separated by 1 to 3 months. However, we regard this as a
valid approximation as
these objects fade very slowly.

Substituting these values 
into equation~\ref{Lha} we obtain a mass-loss rate 
$\dot M$(for $\epsilon_{H\alpha} = 0.1$) $\sim  10^{-2}$ \msun yr$^{-1}$, which
in what follows is  used as reference value. 

The radius of the shock, or inner radius ($R_i$) of the 
unshocked CSM able to produce the narrow 
P-Cygni profile, can be expressed as a function of the local density.
From equations~\ref{mdot} and \ref{Lha} we have:
    \begin{eqnarray}
      \label{ri}
   R_i &  = & 1.54\ET{16}
    \zb{ \frac{ (\dot{M} / 10^{-2} M_\odot yr^{-1}) }{ 
                (v_w / 90 km s^{-1}) } }^{1/2}
    \zp{ \frac{n_i}{10^7 cm^{-3}} }^{-1/2} \,\,\,\,\,\,\,\,\,\,\,\,\,\,\,\, cm   \\
    & = & 1.54\ET{16} 
    \zb{ \frac{ (L_{H\alpha} / 5 \ET{41} erg s^{-1} ) }{ 
                (n_i / 10^7 cm^{-3} ) } }^{1/2}
    \zp{ \frac{v_s}{6.6 \ET{8} cm s^{-1}} }^{-3/2} \,\,\,\, cm  
    \end{eqnarray}
\ni where $n_i$\ is the density of H atoms at $R_i$\ and 
$\rho = 1.4 n_H m_H$\ (accounting for a 10\% He
abundance by number).

\subsection{Narrow P-Cygni lines and the CSM}

We use the luminosities of the emission narrow Balmer lines to
estimate the unshocked mass and  outer radius of the CSM.
These are given by:
\begin{equation}
  \label{lumn}
  L_{Balmer}^{narrow} = \int_{R_i}^{R_o} 4 \pi r^2 n^2 \alpha_{B} h \nu_o dr 
\end{equation}
\ni were $\alpha_{B}$ is the case B recombination coefficient and
we have supposed that the CSM emits isotropically and that it is 
fully ionized.

For this estimation we use either the luminosity of the narrow \ha or \hb lines: The \ha line is more 
intense and it therefore has a better signal-to-noise 
ratio, but in many cases it may be enhanced by collisions and therefore, the 
above relation~(\ref{lumn}) would  no
longer be valid. The \hb line is not affected by collisional effects, but it is 
by reddening.
The ratio \ha/\hb is \si 20, which suggests a high density medium, n \si $\E{8}$ - $\E{10}$,
for an ionization parameter logU = -2 and -4 respectively. 
In what follows we will use the observed values of the 
luminosities in \ha and \hb narrow emission lines. They will gives us an upper and lower
limit respectively to the density and extent of the CSM.

The integration of the above equation using $n = n_i (r/R_i)^{-2}$, yields
\begin{eqnarray}
  L_{Balmer}^{narrow} & = &
 \frac{ \alpha_{B} h \nu_{o} }{ 2 \pi^{1/2} (1.4 m_H)^{3/2} }
 \zp{ \frac{\dot{M}}{v_w} }^{3/2} n_i^{1/2}
 \zp{ 1 - \frac{R_i}{R_o} } \\
   & = & 3.05\ET{37} 
   \alpha_B \nu_{o}
   \zb{ \frac{ (\dot{M} / 10^{-2} M_\odot yr^{-1}) }{ 
                (v_w / 90 km s^{-1}) } }^{3/2}
   \zp{ \frac{n_i}{10^7 cm^{-3}} }^{1/2}
   \zp{ 1 - \frac{R_i}{R_o} }\,\, erg \, s^{-1}
\end{eqnarray}

Using the relationship between \.M and $L^{Broad}_{H\alpha}$ (see equation~\ref{Lha})
one gets:

\begin{eqnarray}
  \label{ggg}
 L_{Balmer}^{narrow} & = &
 \frac{ 4 \alpha_B h \nu_o }{ \pi^{1/2} (1.4 m_H \epsilon_{H\alpha})^{3/2} } 
 \zp{ \frac{L^{Broad}_{H\alpha}}{v_s^3} }^{3/2} n_i^{1/2} 
   \zp{ 1 - \frac{R_i}{R_o} } \\
  &  = & 3.05\ET{37} 
   \alpha_B \nu_{o}
 \zb{ \frac{ (L^{Broad}_{H\alpha} / 5\ET{41} erg s^{-1}) }{ (v_s / 6600 km s^{-1})^3}}^{3/2}
 \zp{ \frac{n_i}{10^7 cm^{-3}} }^{1/2}
   \zp{ 1 - \frac{R_i}{R_o} } \,\, erg \, s^{-1}  
\end{eqnarray}

Substituting the appropriate values for the $L^{narrow}_{H\alpha}$ or $L^{narrow}_{H\beta}$
and $\alpha_B$ and $\nu_o$ we estimated the value of the outer
CSM radius, R$_o$, as a function of the density ($n_i$) at the inner edge ($R_i$) of the CSM.
The observed value of $L_{H\alpha}^{narrow}$ is $4 \ET{39}$ \ergs and of 
$L_{H\beta}^{narrow}$ $2\ET{38}$ erg s$^{-1}$
($H_0 = 75\,\, km s^{-1} Mpc^{-1}$). The true value must be
certainly larger, since the blue side
of the line is in absorption. To account for this and possible
obscuration effects we have multiplied the 
observed emission
luminosities by a factor of 2. The values of $\alpha_B$ taken are:
$\alpha_{H\alpha} = 2.59\ET{-13}$\ cm$^{-3}$ s$^{-1}$\ and 
$\alpha_{H\beta} = 3.07\ET{-14}$\ cm$^{-3}$ s$^{-1}$\ (Osterbrock 1989). 

To explain the observed narrow H$\beta$\
emission, the CSM must have a minimum mass, and this imposes
a natural lower bound on $n_i$:
\begin{equation}
\label{ni}
n_i \ge 3\ET{5} cm^{-3} 
   \zp{ \frac{L_{H\beta}^{narrow}}{10^{38} erg\,s^{-1}} }^2
   \zb{ \frac{ (\dot{M} / 10^{-2} M_\odot yr^{-1}) }{ 
                (v_w / 90 km s^{-1}) } }^{-3}
\end{equation}

\ni and thus, a less dense CSM 
could not produce the observed H$\beta^{narrow}$ line. 
This relationship is valid for an infinite CSM. For
a finite medium, the density must be even larger.

Tables 1 and 2 lists the values of $R_i$, $R_o$, the density ($n_o$)  at $R_o$, as
well as the mass of the
swept up CSM (up to $r = R_i$) and of unshocked CSM mass (from $R_i$ to
$R_o$) for different values of $n_i$. 
We as well list the values of the duration of the progenitor wind, 
t$_{wind}$ = R$_o$ / 90 \kms, and the time for the supernova shock to reach the 
inner radius, t$_{SN}$ = R$_i$ / 10000 \kms, if assumed to travel at 10000 \kms.

\begin{footnotesize}
\begin{table*}
  \caption{CSM values inferred from the H$_{\alpha}^{narrow}$ line}
  \begin{tabular}{|c|c|c|c|c|c|c|c|c|}
    \hline 
   n$_i$   &  R$_i$  &   R$_o$   &  n$_o$  & $\Delta$R  & $M_{shocked}^{CSM}$ & $M_{unshocked}^{CSM}$ & t$_{wind}$ & t$_{SN}$ \\
   \percc  &  cm     &   cm     &  \percc  &  cm       &      \msun      & \msun       & 
 years      &   years \\
    \hline \hline
  $5\ET{7}$  & $6.9\ET{15}$ & $5.9\ET{17}$ & $6.8\ET{3}$ & $5.8\ET{17}$ & 0.24 & 20.4 & 2080 & 0.22  \\
  $10^8$  & $4.9\ET{15}$ & $7.0\ET{15}$ & $4.9\ET{7}$ & $2.1\ET{15}$ & 0.17 & 0.07 &
25 & 0.15 \\
  $10^9$  & $1.5\ET{15}$ & $2.0\ET{15}$ & $6.1\ET{8}$ & $5.0\ET{14}$  & 0.05 & 0.02 &
7   & 0.05 \\
  \hline
  \end{tabular}
  \caption{CSM values inferred from the H$_{\beta}^{narrow}$ line}
  \begin{tabular}{|c|c|c|c|c|c|c|c|c|}
    \hline 
   n$_i$   &  R$_i$  &   R$_o$   &  n$_o$  & $\Delta$R  & $M_{shocked}^{CSM}$ & $M_{unshocked}^{CSM}$ & t$_{wind}$ & t$_{SN}$ \\
   \percc  &  cm     &   cm     &  \percc  &  cm       &      \msun      & \msun       & 
 years      &   years \\
    \hline \hline
  10$^7$  &  $1.5\ET{16}$ & $5.0\ET{16}$ & $9.6\ET{5}$ & $3.5\ET{16}$  &  0.5  & 1.2 
& 175       &  0.5 \\
   $5\ET{7}$  & $6.9\ET{15}$ & $1.0\ET{16}$ & $2.4\ET{7}$ & $3.1\ET{15}$ & 0.2 & 0.1  
& 35          &  0.22 \\
   $10^8$  & $4.9\ET{15}$ & $7.0\ET{15}$ & $6.1\ET{7}$  & $2.1\ET{15}$ & 0.2 & 0.05 
& 22       &  0.15 \\
    $10^9$   & $1.5\ET{15}$ & $2.0\ET{15}$ & $8.7\ET{8}$  & $4.6\ET{14}$ & 0.05 & 0.004
 & 6  &  0.05 \\ 
\hline
  \label{tab:csm}
\end{tabular}
\end{table*}
\end{footnotesize}

Note that the results for both lines agree only for high densities (\si 10$^8$ \percc or
bigger), as we would expect from the high value of $\frac{H\alpha}{H\beta}$ (see previous
section). For such high densities, the extent of the CSM becomes extremely small, and
the density decreases less than one order of magnitude only. This would imply also that
the wind that created such CSM was of very short duration and that the supernova shock
reached it on very short times scales (less than two months).
However, in the derivation of all these parameters, we have assumed that
the narrow Balmer emission lines are due only to recombination effects, and we
did not take into account any collisional effects. 
With the available data we cannot constrain any further these parameters. 
We need more observations to monitor the narrow P-Cygni profile, since its 
disappearance will indicate the end of the dense CSM. 
The numbers presented in tables 1 and 2 must be regarded therefore as an 
indication of the possible range of values. 
Note finally that SN~1997ab was not detected in X-rays by ROSAT 
(A. Fabian, private communication), 
in agreement with the presence of a 
very large column density of neutral material surrounding the SN.
and thus a massive wind.

The rapidly evolving supernova remnant 
is a peculiar emitting source, in the sense that
all of its radiation arises from matter at very high speeds (several thousands of \kms).
This, irrespectively of how dense the circumstellar medium may be, can only be absorbed
by a medium between the source and the observer moving with a similar velocity.
On the other hand, the P-Cygni profile
results from the emission caused by the precursor 
radiation ahead of the blast wave, combined with the absorption 
due to
screening of the radiative supernova remnant shell and the photoionized 
freely expanding ejecta.
From this it follows that a sector of the circumstellar medium, a broad rim, 
overlapping with the section of the high velocity shell that expands near the plane of the
sky and/or with the photoionized freely expanding ejecta also moving near the plane 
of the sky, is the sector that can absorb it and provoke the P-Cygni profiles here
shown. In this respect, the P-Cygni profiles  of SN~1997ab are very different to those
caused by a wind in front of a standing source (e.g. Rublev 1960, Kuan \& Kuhi 1975).
A complete fit to the emission profiles is now in preparation (Cid-Fernandes et al. 1998).

The fact that both the emission and absorption P-Cygni profiles are so similar 
confirms also the fact that the
absorbing CSM rim is in fact very 
broad, and it probably spans from the edge of the 
expanding cooling shell
to the photoionized inner most section of the 
freely expanding ejecta. In this way also, 
there is a central neutral ejecta region that must cause 
the absorption of the redest 
sections of the broad emission lines.

To our knowledge, a narrow P-Cygni profile atop a broad emission
line such as that of SN~1997ab has never been detected in any supernova remnant.
\footnote{Except maybe in SN 1995G, see IAU Circulars number 6138, 6139 and 6140} 
Broader P-Cygni profiles (but still narrower than those seen in typical type II
SN) have been detected in other peculiar SN, such as SN~1979C (Panagia et al. 1980),
SN~1984E (Dopita et al. 1984), SN~1994aj (Benetti et al. 1998) or SN~1996L (Benetti et al. 
1996).
However, the CSM velocities inferred from the widths of all
of these P-Cygni profiles are \si 1000 \kms. If a similar interpretation is given to 
these lines, then the large values of $v_s$ lead for example in the case of 
1994aj to an \.M \si 3.3 10$^{-3}$ \msun yr$^{-1}$ 
and to an $R_i = 1.63\times 10^{15} \frac{1}{\sqrt{n_i/10^7}}$  , 
i.e. \si one order smaller
than the values deduced for 1997ab.
Our result thus broadens the range of values to be expected for the massive winds 
that occur prior to the explosion.

\section*{Acknowledgements} 

I.S. acknowledges the F.P.U. grant from the Ministerio de Educaci\'on 
y Cultura (Spain). R.C.F. work was partially supported by CNPq under 
grant 300867/95-6. The comments suggested by the referee (M. Dopita)
are very appreciated.

\label{lastpage} 
\end{document}